\documentclass[preprintnumbers,amsmath,onecolumn,nofootinbib]{revtex4}

\usepackage{epsfig}
\usepackage{graphicx}
\usepackage{bm}
\usepackage{amsmath}
\usepackage{amssymb}

\newcommand{\si}{\sigma}

\newcommand{\az}{\varphi}
\newcommand{\ro}{\rho}

\newcommand{\oeq}{\begin{equation}}
\newcommand{\ceq}{\end{equation}}
\newcommand{\oeqn}{\begin{eqnarray}}
\newcommand{\ceqn}{\end{eqnarray}}

\renewcommand{\>}{\rangle}
\newcommand{\<}{\langle}
\renewcommand{\(}{\left(}
\renewcommand{\)}{\right)}
\renewcommand{\[}{\left[}
\renewcommand{\]}{\right]}
\renewcommand{\ll}{\left|}
\newcommand{\rl}{\right|}

\newcommand{\stf}{\,\,\,}
\newcommand{\sdf}{\,\,}

\newcommand{\sdb}{\!\!}

\renewcommand{\k}{|}

\newcommand{\kpsi}{|\psi \>}

\newcommand{\bpsi}{\<\psi|}

\newcommand{\oH}{\hat{H}}

\newcommand{\oV}{\hat{V}}

\newcommand{\oro}{\hat{\rho}}
\newcommand{\oh}{\hat{h}}

\newcommand{\ovr}{\hat{\bf r}}

\newcommand{\op}{\hat{p}}

\newcommand{\oad}{\hat{a}^\dagger}
\newcommand{\oa}{\hat{a}}

\newcommand{\oF}{\hat{F}}

\newcommand{\del}{\delta\!}
\newcommand{\dt}{\frac{\partial}{\partial t}}

\renewcommand{\d}{{\mbox d}}

\newcommand{\hb}{\hbar}

\newcommand{\vv}{{\bf v}}

\newcommand{\mL}{{\mathcal{L}}}

\newcommand{\Tr}{\mbox{Tr}}

\begin{document}

\title{Time-Depentent Hartree-Fock description of heavy ions fusion}

\author{C\'edric Simenel and Beno\^{\i}t Avez}

\affiliation{DSM/DAPNIA/SPhN, CEA/SACLAY, F-91191
Gif-sur-Yvette Cedex, France\\
cedric.simenel@cea.fr}

\date{\today}

\begin{abstract}
A microscopic mean-field description of heavy ions fusion
is performed in the framework of the Time-Dependent Hartree-Fock 
(TDHF) theory using a Skyrme interaction with the SLy4$d$ parametrization.
A good agreement with experiments is obtained on the position of the fusion barriers 
for various total masses, mass asymmetries and deformations.
The excitation function of the $^{16}$O+$^{208}$Pb is overestimated by about
16$\%$ above the barrier. The restriction to an independent particles state
in the mean-field dynamics prevents the description of sub-barrier fusion.
Effect of transfer on fusion is discussed. 
\end{abstract}

\maketitle

\section{Introduction}

Description of nuclear reactions is very challenging, 
especially at energies around the fusion barrier
generated by the competition between Coulomb and nuclear
interactions.
It has been established that the structure of the collision partners
may affect strongly the reaction mechanisms in this energy domain, as,
for instance, the fusion cross-sections (for a review, see {e.g.} Ref.~\cite{das98}).

Upcoming exotic beams facilities at energies of a few MeV/u like SPIRAL2 
will allow studies of the interplay between reaction mechanisms 
and exotic nuclear structures such as haloes, neutron skins, high isospins... 
It is therefore recommended to treat both structure and dynamics 
within the same formalism. This is the case of fully microscopic approaches 
such as the Time-Dependent Hartree-Fock (TDHF) theory 
proposed by P. A. M. Dirac in 1930 \cite{dir30}. This time dependent version
of the well known  Hartree-Fock (HF) theory 
gives a self-consistent mean-field description of nuclear dynamics \cite{har28,foc30}. 
The success of the first HF calculations based on the Skyrme interaction \cite{sky56,vau72} led to tremendous activities to describe
 nuclear structure within mean-field based approaches 
(see Ref.~\cite{ben03} for a review). 

Early TDHF calculations has been done with the seek for 
a description of the dynamics of nuclei as good as their static 
properties \cite{eng75,bon76,bon78,flo78,kri78,dav78,dev81,kri81,neg82}.  
They used various symmetries and 
simplified Skyrme interactions to reduce the computational 
time. The increase of computational power allowed 
realistic TDHF calculations of nuclear collisions in 3 dimensions 
with full Skyrme interactions in the last ten 
years \cite{kim97,sim01,nak05,umar,mar06,guo07,sim07}.

In this paper, we present a TDHF study of nuclear fusion.
In a first part, we recall the formalism and detail the calculation.
In a second part, we calculate fusion barriers for several systems with various
total masses and mass asymmetries and compare them to experimental data. 
We also consider the case where one of the collision partners is prolately deformed.
Finally, we calculate the excitation function for the total fusion cross section 
of the system $^{16}$O+$^{208}$Pb and compare to experiments
 before to conclude.

\section{The Time-Dependent Hartree-Fock Approach}

\subsection{Formalism}

Let us first recall some aspects of the  TDHF theory.
In a non relativistic microscopic approach, the system 
 is described by a N-particles state $\kpsi$  
which is solution of the Schroedinger equation
\oeq
i\hb \sdf \dt \kpsi = \oH \sdf \kpsi 
\label{eq:schroed}
\ceq
with the microscopic Hamiltonian\footnote{We consider only two-body interactions.}
\oeq
\oH = \sum_{i=1}^N \sdf \frac{\op{(i)}^2}{2m} + \sum_{i>j=1}^N \sdf \hat{v}{(i,j)}.
\label{eq:oH1}
\ceq
The state $\kpsi$ contains all the information on the system, 
which is more than what we really need for a good 
description of the dynamics. We often need only
 expectation values of one-body observables, such as 
 the position of the fragments, their shapes and particle numbers.
These quantities  are determined from the one-body density matrix $\ro$
with elements $\ro_{ij} = \bpsi \oad_j \oa_i \kpsi$. 
The expectation value of a one-body observable $\oF = \sum_{ij} f_{ij}\, \oad_i \oa_j$
is then given by $\bpsi \oF \kpsi = \Tr \(\ro f\).$

The first step toward the TDHF theory is to restrict the description 
to one-body observables, and to seek for an equation giving the evolution
of $\ro$. Starting from Eq.~(\ref{eq:schroed}) and 
using the Bogolyubov-Born-Green-Kirkwood-Yvon (BBGKY) 
hierarchy \cite{bog46,bor46,kir46},
we can show that the one-body density matrix follows \cite{lac03}
\oeq
i\, \hb \sdf \dt \sdf {\ro} =  \[\sdf h[\ro] \, ,\, \ro\sdf \] +\Tr_2 \[ \sdf v(1,2) \, , \sdf C(1,2) \sdf  \].
 \label{eq:TDDM}
\ceq
where $h[\ro]$  is the HF single-particle Hamiltonian with
matrix elements $h_{ij}=\<i|\, \oh[\ro]\,|j\>=\frac{\del}{\del \ro_{ji}}\sdf \bpsi \oH \kpsi$
and $C$ is  the correlated part of the two-body density matrix. 

Eq.~(\ref{eq:TDDM}) is exact but has two unknown quantities: $\ro$ and $C$. 
The second step toward the TDHF equation is to
neglect the second term of the right hand side in Eq.~(\ref{eq:TDDM}).
This can be done in two alternative ways:
\begin{itemize}
\item The correlation $C$ vanishes if we impose $\kpsi$ 
to be an independent particles state at any time. 
The variational principle $\delta \[\int \sdb \d t \stf \bpsi \oH - i\hb \dt \kpsi\] = 0$,
which is equivalent to Eq.~(\ref{eq:schroed}),
is then solved in the subspace of Slater determinants.
\item The truncation of the BBGKY hierarchy can also be done 
by neglecting the residual interaction
$\oV_{res} = \oH-\sum_{i=1}^N\oh[\ro](i)$.
This is a mean-field approximation because the Hamiltonian
is approximated by a one-body operator 
$\oH \simeq \sum_{ij} h_{ij}\, \oad_i \oa_j $.
In this case, a system described by a Slater determinant at an initial time 
will be an independent particles state at any time. 
\end{itemize}
We finally get the TDHF equation
$ i\hb \sdf \dt \, \ro = \[\,h[\ro]\, , \, \ro \,\]$
where $\ro$ is now the one-body density matrix of
an independent particles state. The operator 
associated to $\ro$ acts in the Hilbert space 
of single-particle states. It is  written 
$\oro=\sum_{i=1}^{N}\ll \az_{i}\> \<\az_{i}\rl$
where $\k \az_i \>$ denotes an occupied single-particle state.

The TDHF theory neglects the pairing correlations which are contained 
in $C$. In fact, TDHF describes the evolution of occupied single-particle 
wave functions in the mean field generated by all the particles and assures an
exact treatment of the Pauli principle during the dynamics.

\subsection{Practical aspects}

The advantage of TDHF is that it treats static
properties {\it and} dynamics of nuclei within the same
formalism and the same interaction. The initial state is obtained
through static HF calculations which 
reproduce well  nuclear binding energies and
deformations. TDHF can be used in two ways for nuclear
reactions:
\begin{itemize}
\item A single nucleus is evolved in an external field \cite{eng75}, 
simulating for instance the Coulomb field of the collision partner \cite{sim04}.
\item The evolution of two nuclei, initially with a zero overlap,
is represented by a single Slater determinant \cite{bon76,neg82}.
\end{itemize}
The first case 
is well suited for inelastic scattering, like Coulomb 
excitation of vibrational and
rotational states. The second case is used for more violent
collisions like fusion reactions. In the latter, 
the lack of a collision term might be a drawback.
At low energy, however,  fusion is driven by one-body
dissipation because the Pauli blocking prevents 
nucleon-nucleon collisions.  Fusion occurs by
transferring relative motion  into internal excitation via
one-body mechanisms well treated by TDHF.

Another important advantage of TDHF for near-barrier reaction studies 
is that it contains all  types of couplings between the relative
motion and internal degrees of freedom whereas in coupled
channels calculations one has to include them explicitly
according to physical intuition. The symmetries corresponding 
to the internal degrees of freedom of interest have to be relaxed in TDHF. 
However, it gives only classical
trajectories for the time-evolution  and
 expectation values of one-body observables. In particular,
it does not include tunneling of the global wave function.

We use the TDHF3D code built by P. Bonche and coworkers with the SLy4$d$
 Skyrme parametrization \cite{kim97} which is a variant 
 of the SLy4 one \cite{cha98} specifically designed for TDHF 
 calculations\footnote{The calculations are performed 
 in the laboratory frame and not in the intrinsic frame. 
 It is then necessary to remove the center of mass correction 
 in the fitting procedure of the interaction and in the initial HF calculations 
 of the collision partners. The "$d$" in SLy4$d$ stands for {\it dynamics}.}.
This code has a plane of symmetry (the collision plane).
It uses the Skyrme energy functional expressed in Eq.~(A.2) of Ref.~\cite{bon87} 
where the tensor coupling between spin and gradient has been neglected.
 The step size of the network is 0.8~fm and the step
time 0.45~fm/c.

A TDHF calculation of two colliding nuclei is performed assuming that 
the two collision partners are initially at a distance
$D_0$ in their HF ground state. This distance has to be big enough 
to allow Coulomb excitation in the entrance channel 
(polarization, vibration, rotation...).
This initial distance is chosen to be $D_0=44.8$~fm. 
We assume that before to reach this distance,
the nuclei followed a Rutherford trajectory, which determines their 
initial velocities $\vv_1$ and $\vv_2$. 
The Galilean transformation
$
\oro_i(t=0) = e^{i\,m\,\vv_i\cdot \ovr} \sdf \oro_{HF_i} \sdf e^{-i\,m\,\vv_i\cdot \ovr}
$
applied on the HF density matrix of the nucleus $i$ ($i=1$ or 2) 
put it into motion with the velocity $\vv_i$ \cite{tho62}.

\section{Fusion Barriers}

Let us consider the simple case of fusion barriers.
They are classically defined as the energy threshold above which  
fusion occurs for a head-on collision. Experimentally, 
the average position of the barrier can be approximated by the centroid 
of the so-called barrier distribution $D_B(E) = \frac{\d^2}{\d E^2} \sdf \(\sigma_{fus}(E)\sdf E\)$ where $\sigma_{fus}$ is the fusion cross section \cite{row91}.
In the special case of a single barrier for a classical system, the barrier distribution
is a Dirac distribution $D_B \sim \delta (E-B)$, whereas a width is generated
by tunneling for a quantum system. Several barriers may be generated by
 the coupling between internal degrees of freedom and the relative motion \cite{das98}.

To determine the fusion barrier from TDHF, we consider head-on collisions
at various energies. The barrier is then located between the highest energy 
for which there is no fusion and the lowest one for which fusion occurs.
The average barrier is first studied for spherical nuclei.
Then the case of deformed nuclei is considered.

\subsection{Spherical nuclei}

\begin{figure}[th]
\centerline{\psfig{file=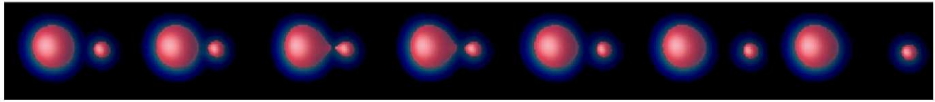
,width=12.5cm}}
\vspace*{8pt}
\caption{Densities associated to a $^{16}$O+$^{208}$Pb 
central collision at a center of mass energy $E=74.44$ MeV.
The surface corresponds to an isodensity at half the saturation density. 
Each plot is separated by 135 fm/c.}
\label{fig:74_44}
\end{figure}

\begin{figure}[th]
\centerline{\psfig{file=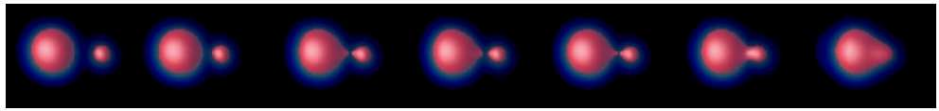
,width=12.5cm}}
\vspace*{8pt}
\caption{Same as Fig. \ref{fig:74_44} at $E=74.45$ MeV.}
\label{fig:74_45}
\end{figure}

Fig. \ref{fig:74_44} shows the density plot for a $^{16}$O+$^{208}$Pb 
central collision at a center of mass energy $E=74.44$ MeV. 
After a neck formation, the system separates into two fragments. 
Fig. \ref{fig:74_45} shows the same reaction at $E=74.45$ MeV.
We see that adding 10~keV is enough to fuse. We deduce 
the fusion barrier $V_B^{TDHF}= 74.445\pm0.005$~MeV.
This value is in excellent agreement with 
the experimental one
$V_B^{exp.}\simeq 74.5$ MeV \cite{mor99}. 
It is interesting to note that, if we assume frozen HF densities 
of the collision partners obtained with the same interaction, 
we get a barrier $V_B^{frozen}=76.55$ MeV at a radius 
$R_B^{frozen}=11.73$ fm. These values are close to the Bass barrier
$V_B^{Bass}=77.10$ MeV at $R_B^{Bass}=11.42$ fm \cite{bas77,bas80},
but overestimate the experimental fusion barrier.
We conclude that the TDHF calculations contain dynamical
effects that reduce the barrier by $\sim 2$ MeV
 as compared to the frozen approximation.
It has been suggested that transfer may affect the barrier 
for this system \cite{mor99}. Indeed, we see in Fig. \ref{fig:74_45} 
that the two fragments are linked 
by a neck and form a di-nuclear system during $\sim400$~fm/c. 
It is enough time for the nuclei to transfer nucleons,
leading to a dynamical evolution of the barrier.  

To get a deeper insight into this transfer preceding fusion, 
we focus on the case just below the barrier (Fig. \ref{fig:74_44}).
Here, almost two protons and no neutron, in average, 
have been transfered from the $^{16}$O to the $^{208}$Pb. 
The two-protons transfer from the light to the heavy nucleus 
 is then expected to be an important channel at the barrier. 
 This is consistent with experimental observations of 
 relatively high Carbon production cross sections, 
 of the same order of the Nitrogen ones, in the exit channel
 of $^{16}$O+$^{208}$Pb at the barrier \cite{vid77,vul86}.

\begin{figure}[th]
\centerline{\psfig{file=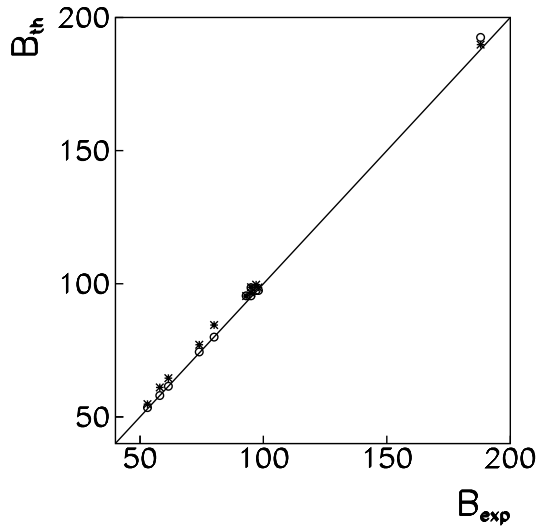,width=7cm}}
\vspace*{8pt}
\caption{Theoretical fusion barriers from TDHF calculations (circles) 
and the Bass barriers (stars) as function of the experimental values
(from  barrier distributions centroid).}
\label{fig:barriers}
\end{figure}

Finally, Fig. \ref{fig:barriers} shows a comparison between experimental 
fusion barriers and those from TDHF calculations
for systems with various total masses and mass asymmetries.
The lowest barrier is for $^{40}$Ca+$^{40}$Ca and the highest one for 
$^{48}$Ti+$^{208}$Pb.   
The  agreement is of the same order 
than with the Bass barrier \cite{bas77,bas80}.
Remembering that TDHF has no adjustable parameter on reaction mechanisms, 
we conclude that one can use it for fusion barriers prediction with confidence.

\subsection{Effect of deformation on fusion}

We now consider collisions of a spherical nucleus on a deformed one. 
In such a case, the barrier depends on the orientation of the deformed
nucleus at the touching point, leading to a wider barrier distribution 
than the single barrier case \cite{das98}. 

\subsubsection{Heavy deformed + light spherical nuclei}

\begin{figure}[th]
\centerline{\psfig{file=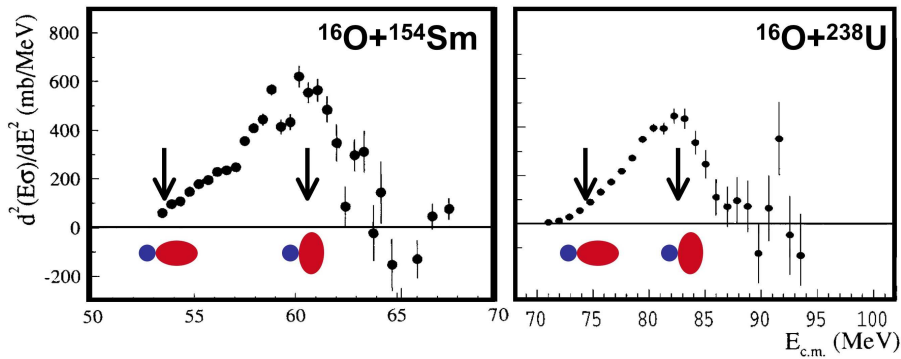,width=12cm}}
\vspace*{8pt}
\caption{Experimental barrier distributions for $^{16}$O+$^{154}$Sm (left) 
and $^{16}$O+$^{238}$U (right).
The arrows indicate the barriers obtained from TDHF calculations for two 
extreme configurations where the collision and deformation axis are parallel
(lowest barriers) and perpendicular (highest barriers).}
\label{fig:def}
\end{figure}

Fig. \ref{fig:def} shows barrier distributions
for  collisions with a light spherical projectile ($^{16}$O) on 
heavy prolately deformed targets $^{154}$Sm (left)  
and $^{238}$U (right) \cite{lei95,hin96}.
Their width are $\sim 7-10$ MeV and cannot be explained by
tunneling alone. The latter adds a width of only 2-3 MeV \cite{row91}. 
Such barrier distributions are usually well reproduced in the 
framework of coupled channel calculations \cite{das98}.
In addition, microscopic theories like TDHF can help to understand 
the physics process generating these couplings.

The barriers predicted by the TDHF calculations for these systems are 
also shown (arrows) on Fig. \ref{fig:def} for two 
extreme configurations of central collisions where the
collision and deformation axis are either parallel or perpendicular. 
In the parallel configuration, the Coulomb repulsion is smaller
and so is the resulting barrier.
All the intermediate orientations at the touching point 
give fusion barriers between the parallel and perpendicular configurations ones.
In addition to a good reproduction of the centroid,
the TDHF calculations also reproduce the width of the barrier distributions
generated by a static deformation of the target without any adjustment of parameters. 

The shape of these distributions is due to a prolate deformation of the target:
the distribution is more peaked at higher energies.
The configuration where the
 collision and deformation axis are parallel 
corresponds to only one possible orientation, 
whereas the perpendicular configuration can be reached by 
any orientation for which the deformation axis is 
contained in the plane perpendicular to the collision axis. 
If one assumes an isotropic distribution of the orientations at the touching point, 
which is a reasonable approximation when the spherical nucleus is light and
the deformed one heavy \cite{sim04}, then the perpendicular configuration 
is more probable than the parallel one and the barrier distribution is peaked
at high energies. 
One gets the opposite with an oblate nucleus instead of a prolate one.

\subsubsection{light deformed + heavy spherical nuclei}

The case of a light deformed projectile on a heavy spherical target
has been investigated theoretically in Ref.~\cite{sim04} both within the TDHF 
and the coupled channel frameworks with the code CCFULL \cite{hag99}.
For such systems, the barrier distribution 
gets affected due to reorientation of the deformed nucleus
in the Coulomb field of the target. This breaks the isotropy
of the orientation axis distribution and results to 
a fusion hindrance at low energies.
Experimental evidences of this effect have been reported recently \cite{nay07}.
The reorientation is proportional to
$A_{spherical}/A_{total}$  and then  does not affect the systems studied in 
Fig. \ref{fig:def}\footnote{Though it is a Coulomb effect, 
the reorientation does not depend on the charges of the nuclei 
but on their masses as we can see in Eq.~(5) of Ref.~\cite{sim04}.}.

\section{Excitation Function of $^{16}$O+$^{208}$Pb}

We now focus on fusion cross sections given by
\oeq
\si_{fus} (E) = \frac{\pi \hbar^2}{2\mu E} \sdf 
\sum_{l=0}^{\infty} \sdf (2l+1) \sdf P_{fus}(E,l)
\ceq
where $\mu$ is the reduced mass of the system and
$P_{fus}(E,l)$ is the fusion probability at a center of mass energy $E$ 
and an angular momentum $\sqrt{l(l+1)}\hbar$. 
The restriction to an independent particles state, as in TDHF, 
leads to $P_{fus}=1$ for $l\le l_{max}$ and 0 for $l>l_{max}$. 
We then get the so-called "quantum sharp cut-off formula" \cite{bla54}
\oeq
\si_{fus} (E) = \frac{\pi \hbar^2}{2\mu E} \sdf (l_{max}+1)^2.
\ceq
To avoid discontinuities introduced by the cut-off and 
the integer nature of $l_{max}$,
we approximate $(l_{max}+1)\hbar$ by its classical equivalent $\mL_c$.
The latter is the threshold of the classical angular momentum 
$\mL = \sqrt{2\mu E}\, b$, $b$ being the impact parameter, 
below which fusion occurs \cite{bas80}.
This approximation is justified by the fact that both $(l_{max}+1)^2$ and
$\mL_c^2/\hbar^2$ are greater than $l_{max}(l_{max}+1)$ and smaller
than $(l_{max}+1)(l_{max}+2)$. We finally get the classical expression 
for the fusion cross section $\si_{fus}(E) \simeq \pi \mL_c^2/2\mu E$.

\begin{figure}[th]
\centerline{\psfig{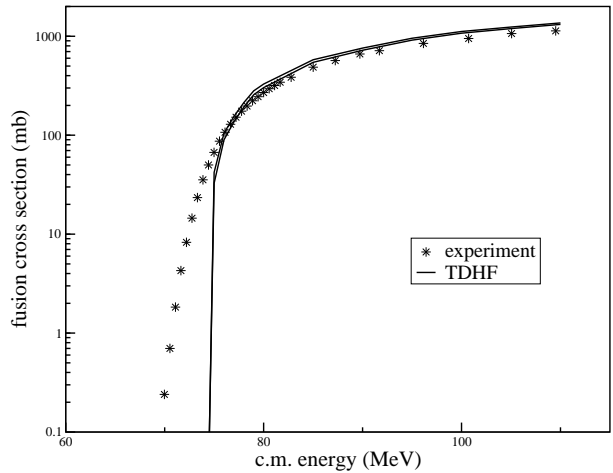}}
\vspace*{8pt}
\caption{Experimental fusion excitation 
function (stars) for $^{16}$O+$^{208}$Pb. The lines denote the upper and lower
limits for the fusion cross sections obtained from TDHF calculations.}
\label{fig:fus}
\end{figure}

Figure \ref{fig:fus} shows the excitation function obtained for the 
$^{16}$O+$^{208}$Pb system in comparison to experimental data \cite{mor99}. 
There is a good agreement above the barrier, though
the fusion cross sections are overestimated by about 16 $\%$.
However, the fusion cross section vanishes below the barrier, following 
a classical behavior. This drawback of TDHF is well known and is 
due to the restriction to a single independent particles state. 
Indeed, to get a fusion probability
between 0 and 1, we need at least two Slater determinants: 
one describing the two well separated fragments after the collision
when fusion does not occur
and one describing the fused system. 
It is necessary to go beyond TDHF to treat a sum of Slater determinants
and then to describe 
sub-barrier fusion due to tunneling effects.

\section{Conclusions and Perspectives}
We presented some applications of the Time-Dependent Hartree-Fock
theory to nuclear fusion. The only phenomenological input 
is the set of parameters of the SLy4$d$ force which have not been
adjusted on any reaction mechanism like cross sections or fusion barriers
for instance. Despite this, the agreement between the TDHF calculations 
and the experimental fusion barriers is excellent for a wide range 
of projectiles and targets. The width of fusion barrier distributions
generated by a static deformation of the target is also well reproduced. 

The TDHF calculations overestimate
 the fusion cross sections for the system $^{16}$O+$^{208}$Pb 
 above the barrier by about 16$\%$.
Below the barrier, the fusion cross section vanishes in TDHF calculations.
The sub-barrier fusion due to quantum tunneling of the many-body wave function
 is not present in TDHF. 
 This is due to the restriction to a single independent particles state. 

Though the fusion cross section has a classical behavior, 
 the quantum nature of the single-particle 
wave functions is well treated. Thus, it would be interesting to study 
 the transfer of nucleons from one nucleus to the other within 
 TDHF\footnote{We have to distinguish between the transfer of typically
 one or two nucleons and more violent collisions like deep-inelastic reactions.
 In the latter, the width of the particle number distribution is known to be 
 underestimated with a single Slater determinant \cite{das79}.}.
An illustration of such transfer is seen in the TDHF calculation
of the $^{16}$O+$^{208}$Pb just below the barrier. 
 
\section*{Acknowledgements}

We thank P. Bonche who provided his TDHF code and initiated this work. 
We are also grateful to D. Lacroix, M. Bender, K. Bennaceur, Ph. Chomaz
and T. Duguet  for fruitful discussions. 
The calculations have been performed in the 
Centre de Calcul Recherche et Technologie of the Commisariat \`a l'\'Energie Atomique.


\begin{thebibliography}{0}

\bibitem{das98}
M. Dasgupta, D. J. Hinde, N. Rowley and A. M. Stefanini, {\it Annu. Rev. Nucl. Part. Sci.} {\bf 48} (1998) 401 {\it and ref. therein}.

\bibitem{dir30} P. A. M. Dirac, {\it Proc. Camb. Phil. Soc.} {\bf 26} (1930) 376.

\bibitem{har28}  D. R. Hartree, {\it Proc. Camb. Phil. Soc.} {\bf 24} (1928) 89.

\bibitem{foc30}  V. A. Fock, {\it Z. Phys.}  {\bf 61} (1930) 126.

\bibitem{sky56}  T. Skyrme, {\it Phil. Mag.} {\bf 1} (1956) 1043.

\bibitem{vau72}  D. Vautherin and D. M. Brink, {\it Phys. Rev.}  {\bf C5} (1972) 626.

\bibitem{ben03} M. Bender, P.-H. Heenen and P.-G. Reinhard,  
{\it Rev. Mod. Phys.} {\bf 75} (2003) 121.

\bibitem{eng75} Y. M. Engel {\it et al.}, 
{\it Nucl. Phys.} {\bf A249} (1975) 215.

\bibitem{bon76}  P. Bonche, S. Koonin and J. W. Negele,
{\it Phys. Rev.} {\bf C13} (1976) 1226.

\bibitem{bon78} P. Bonche, B. Grammaticos and S. Koonin,
{\it Phys. Rev.} {\bf  C17} (1978) 1700.

\bibitem{flo78} H. Flocard, S. E. Koonin and M. S. Weiss,
{\it Phys. Rev.} {\bf C17} (1978) 1682.

\bibitem{kri78} S. J. Krieger and K. T. R. Davis, {\it Phys. Rev.} {\bf C18} (1978) 2567.

\bibitem{dav78}
K. T. R. Davies, V. Maruhn-Rezwani, S. E. Koonin and J. W. Negele,
{\it Phys. Rev. Lett.}  {\bf 41} (1978) 632. 

\bibitem{dev81} K. R. S. Devi, A. K. Dhar and M. R. Strayer,
{\it Phys. Rev.} {\bf C23} (1981) 2062.

\bibitem{kri81} S. J. Krieger and M. S. Weiss,
{\it Phys. Rev.}  {\bf C24} (1981) 928. 

\bibitem{neg82}  J. W. Negele, {\it Rev. Mod. Phys.} {\bf 54} (1982) 913.

\bibitem{kim97}  K.-H. Kim, T. Otsuka and P. Bonche, {\it J. Phys.}  {\bf G23} (1997) 1267.

\bibitem{sim01} C. Simenel, Ph. Chomaz and G. de France, {\it Phys. Rev. Lett.} {\bf 86} (2001) 2971.

\bibitem{nak05} T. Nakatsukasa and K. Yabana, {\it Phys. Rev.} {\bf C71} (2005), 024301.

\bibitem{umar}
A. S. Umar and V. E. Oberacker, 
{\it Phys. Rev.} {\bf C73} (2006) 054607; 
{\it Phys. Rev.} {\bf C74} (2006) 021601(R); 
{\it Phys. Rev.}  {\bf C74} (2006) 024606;
{\it Phys. Rev.}  {\bf C74} (2006) 061601;
{\it Phys. Rev.}  {\bf C76} (2007) 014614.

\bibitem{mar06}
J. A. Maruhn, P.-G. Reinhard, P. D. Stevenson and M. R. Strayer, {Phys. Rev.} {\bf C74} (2006) 027601.



\bibitem{guo07}
L. Guo, J. A. Maruhn and P.-G. Reinhard, {\it Phys. Rev.} {\bf C76} (2007) 014601.

\bibitem{sim07} C. Simenel, Ph. Chomaz and G. de France, {\it Phys. Rev.} {\bf C76} (2007) 024609.

\bibitem{bog46} N. N. Bogolyubov, {\it J. Phys. (URSS)} {\bf 10} (1946) 256.

\bibitem{bor46} H. Born and H. S. Green, {\it Proc. Roy. Soc.} {\bf A188} (1946) 10.

\bibitem{kir46} J. G.Kirwood, {\it J. Chem. Phys.} {\bf 14} (1946) 180.
 
 \bibitem{lac03} D. Lacroix, S. Ayik and Ph. Chomaz, {Prog. Part. Nucl. Phys.} {\bf 52} (2004) 497.

\bibitem{sim04} C. Simenel, Ph. Chomaz and G. de France, {\it Phys. Rev. Lett.} {\bf 93} (2004) 102701.

\bibitem{cha98}
E. Chabanat {\it et al.}, 
{\it Nucl. Phys.} {\bf A635}  (1998) 231.

\bibitem{bon87} P. Bonche, H. Flocard and P.-H. Heenen, {\it Nucl. Phys.} {\bf A467} (1987) 115.

\bibitem{tho62} D. J. Thouless and J. G. Valatin, {\it Nucl. Phys.} {\bf 31} (1962) 211.

\bibitem{row91} N. Rowley, G. R. Satchler and P. H. Stelson,
{\it Phys. Lett.} {\bf B254} (1991) 25.

\bibitem{mor99} C. R. Morton {\it et al.,} {\it Phys. Rev.} {\bf C60} (1999) 044608.

\bibitem{bas77} R. Bass, {\it Phys. Rev. Lett.} {\bf 39} (1977) 265.

\bibitem{bas80} R. Bass, {\it Nuclear Reactions} (Springer-Verlag, Berlin, 1980).

\bibitem{vid77} F. Videb\ae k {\it et al.}, {\it Phys. Rev.} {\bf C15} (1977) 954.

\bibitem{vul86} E. Vulgaris, L. Grodzins, S. G. Steadman and R. Ledoux,
{\it Phys. Rev.} {\bf C33} (1986) 2017.
 
\bibitem{lei95} J. R. Leigh {\it et al.,} {\it Phys. Rev.} {\bf C52} (1995) 3151.

\bibitem{hin96} D. J. Hinde {\it et al.,} {\it Phys. Rev.} {\bf C53} (1996) 1290.

\bibitem{hag99}  K. Hagino, N. Rowley and A. T. Kruppa,
{\it Comp. Phys. Com.} {\bf 123} (1999) 143.

\bibitem{nay07} B. K. Nayak {\it et al.,} {\it Phys. Rev.} {\bf C75} (2007) 054615.

\bibitem{bla54} J. S. Blair, {\it Phys. Rev.} {\bf 95} (1954) 1218.

\bibitem{das79} C. H. Dasso, T. D\o ssing and H. C. Pauli, {\it Z. Phys.} {\bf A289} 
(1979) 395.

\end{thebibliography}
\end{document}